\documentclass[12pt]{article}
\def\be {\begin{equation}}
\def\ee {\end{equation}}
\def\ba {\begin{eqnarray}}
\def\ea {\end{eqnarray}}

\def\lb {\label}
\def\order{{\cal {O}}}
\def\e  {\epsilon}

\def\th {\theta}

\def\bi {\begin{itemize}}
\def\ei {\end{itemize}}
\begin{document}
\def\bea{\begin{eqnarray}}
\def\eea{\end{eqnarray}}

\title{Microscopic entropy of the charged BTZ black hole
}

\author{ Mariano Cadoni\footnote{email: mariano.cadoni@ca.infn.it }\,\,$^1$,
Maurizio Melis\footnote{email: maurizio.melis@ca.infn.it  }\,\,$^{1}$
and Mohammad R. Setare \footnote{email: rezakord@ipm.ir}\,\,$^{2}$
\\
{$^1$\it \small Dipartimento di Fisica, Universit\`a di Cagliari and
INFN, Sezione di Cagliari }\\
{\small Cittadella Universitaria, 09042 Monserrato, Italy}\\
{$^2$ \it \small Department of Science,  Payame Noor University.
Bijar, Iran}\\}
\vfill

\maketitle

{\bf Abstract.}
The charged BTZ black hole is characterized by a power-law curvature
singularity  generated by the electric charge of the hole.
The curvature singularity
produces $\ln r$ terms in the asymptotic expansion of the
gravitational field and divergent contributions to the
boundary terms.
We  show that these boundary deformations can be
generated by the action of the conformal group in two dimensions and that an
appropriate renormalization procedure allows for the definition of
finite boundary charges.
 In the semiclassical regime the central charge of the dual CFT 
 turns out to be that calculated by Brown and Henneaux, whereas the charge
associated with time translation is given by the renormalized black 
hole  mass.
We  then show that  the Cardy formula reproduces exactly the
Bekenstein-Hawking entropy of the charged BTZ black hole.

\section{Introduction}
The discovery of the existence of black hole solutions in three
spacetime dimensions by Ba\~nados, Teitelboim and Zanelli (BTZ)
\cite{Banados:1992wn,Banados:1992gq} (for a review see Ref.
\cite{Carlip:1995qv})
represents one of the main recent  advances for low-dimensional
gravity theories. 
Owing to its simplicity and to the fact that it can be formulated as
a Chern-Simon theory, 3D gravity has  become paradigmatic for
understanding general features of gravity, and in particular its
relationship with gauge field theories,  in any spacetime dimensions.

The realization of the existence of three-dimensional (3D) black holes not only deepened our understanding of
3D gravity but also became a  central key for recent developments in
gravity, gauge and string theory.

In this context an important role is played by the notion of 
asymptotic symmetry.
This notion was  applied with
success  some time ago  to  asymptotically 3D anti-de Sitter 
(AdS$_{3}$) spacetimes, to show that the
asymptotic symmetry  group (ASG) of AdS$_{3}$ is the
conformal group in two dimensions \cite{Brown:1986nw}. This fact
represents the first evidence of the existence of an anti-de 
Sitter/conformal field theory (AdS/CFT)
correspondence and was later used by Strominger to explain the
Bekenstein-Hawking entropy of the BTZ black hole in terms of the
degeneracy of states of the boundary CFT generated by the asymptotic
metric deformations \cite{Strominger:1997eq}.
Moreover, the Chern-Simon formulation of 3D gravity
allowed to give a nice physical interpretation of the degrees of
freedom whose degeneracy should account for the Bekenstein-Hawking
entropy of the BTZ black hole \cite{Carlip:1998qw,Carlip:1998uc,Carlip:2005zn}.

Nowadays, the best-known  example of the AdS/CFT correspondence
\cite{Maldacena:1997re,Witten:1998qj} is
represented by bulk 3D gravity whose dual is a  two-dimensional (2D)
conformal field theory (CFT). The BTZ black hole fits nicely  in the
AdS/CFT framework and  can be interpreted as  excitation of the
AdS$_{3}$ background, which is dual to thermal excitations
of the boundary CFT.
The BTZ black
hole continues to play a key role in  recent investigations
aiming to improve our understanding of 3D  gravity  and of  general
feature of the gravitational interaction \cite{Witten:2007kt}.

A characterizing feature of the  BTZ black hole (at least in its
uncharged form)  is the absence of  curvature
singularities. The
scalar curvature is well-behaved (and constant) throughout the whole 3D
spacetime.
This feature is shared by other low-dimensional examples such as 2D
AdS black holes (see e.g. Ref. \cite{Cadoni:1994uf}), for which also the
microscopic entropy could be calculated \cite{Cadoni:1998sg,Cadoni:1999ja} using the method
proposed in Ref. \cite{Strominger:1997eq}.

On the other hand the absence of curvature singularities
makes the BTZ black hole very different from its higher dimensional cousins
such as the 4D Schwarzschild black hole. Obviously, this difference
represents a loss of the ``paradigmatic power'', which the BTZ black
hole has in the context of the theories of gravity.
One can  try to
consider low-dimensional black holes  with curvature singularities
generated by matter sources. But, in general the presence of these
sources  generates  a gravitational field which  asymptotically
falls off less rapidly then the AdS term
producing divergent boundary terms \cite{Cadoni:2001ew}.

In this paper we  consider the alternative case in which the curvature
singularity is not generated by mass sources but by charges of the
matter fields. Because  matter fields fall off more
rapidly then the gravitational field, we expect the divergent
boundary contributions to be much milder and removable by an 
appropriate renormalization procedure.

An example of this behavior, which we discuss in detail in this
paper, is the electrically charged BTZ black hole.
It is characterized by a power-law curvature singularity  generated
by the electric charge of the hole. The curvature singularity
generates $\ln r$ terms in the asymptotic expansion of the
gravitational field, which give divergent contributions to the
boundary terms. We will show that these boundary deformations can be
generated by the action of the  conformal group and that an
appropriate renormalization procedure allows for the definition of
finite boundary charges. The central charge of the dual CFT turns out
to be the same as that calculated in Ref. \cite{Brown:1986nw}, whereas the charge
associated with time translation is given by the renormalized mass.
We  then show that  the Cardy formula reproduces exactly the
Bekenstein-Hawking entropy of the charged BTZ black hole.

\section{The charged BTZ black hole }
The BTZ black hole solutions 
\cite{Banados:1992wn,Banados:1992gq} in $(2+1)$ spacetime dimensions are derived
from a three-dimensional theory of gravity
\be I=\frac{1}{16 \pi G}\int d^{3}x
\sqrt{-g}\,( R+2\Lambda)\label{ac1} \ee
where $G$ is the 3D Newton constant
and $\Lambda=\frac{1}{l^2}>0$ is
the cosmological constant. We are using units where $G$
and $l$ have both the dimension of a length \footnote{Notice that
often in the literature units are chosen such that $G$ is dimensionless, $8G=1$.}.
\par\noindent
The corresponding line element in Schwarzschild coordinates is
\be ds^2 =- f(r)dt^2
+ f^{-1}dr^{2}¥+r^2\left(d\theta -\frac{4GJ}{r^2}dt\right)^2,
\label{metric}\ee
with metric function: \be
f(r)=-8GM+\frac{r^2}{l^2} +\frac{16 G^{2}¥J^2}{
r^2},\label{metric2}
 \ee where $M$ is the Arnowitt-Deser-Misner (ADM) mass,
$J$ the angular momentum (spin)
 of the BTZ black hole and $-\infty<t<+\infty$, $0\leq r<+\infty$,
 $0\leq \theta <2\pi$.
The outer and inner horizons, i.e. $r_{+}$ (henceforth simply black
hole horizon) and $r_{-}$ respectively, concerning the positive mass
black hole spectrum with spin ($J\neq 0$) of the line element,
(\ref{metric}) are given by  \be
r^{2}_{\pm}={4Gl^2}\left(M\pm\sqrt{M^2 -
\displaystyle{\frac{J^2}{l^2}} }\right). \label{horizon1} \ee In
addition to the BTZ solutions described above, it was also shown in
\cite{Banados:1992wn,Martinez:1999qi} that charged black hole solutions similar to
(\ref{metric}) exist. These are solutions following from the action
\cite{Martinez:1999qi,Achucarro:1993fd}
\be I=\frac{1}{16\pi G}\int d^{3}x \sqrt{-g}\,
(R+2\Lambda-4\pi G F_{\mu\nu}F^{\mu\nu}) \label{ac2}. \ee
The Einstein
equations are given by
\be \label{ein}G_{\mu\nu}-\Lambda
g_{\mu\nu}=8\pi G T_{\mu\nu}, \ee
where $T_{\mu\nu}$ is the
energy-momentum tensor of the electromagnetic (EM) field: \be \label{tmn}
T_{\mu\nu}=F_{\mu\rho}F_{\nu\sigma}g^{\rho\sigma}-\frac{1}{4}g_{\mu\nu}F^2.
\ee
 Electrically charged black hole solutions of the equations (\ref{ein})
 take the
form (\ref{metric}), but with
\be \label{charged}
f(r)=-8GM+\frac{r^2}{l^2}
+\frac{16 G^{2}¥J^2}{r^2}-8\pi GQ^2 \ln (\frac{r}{l}), \hspace{0.5cm}
\ee
whereas the $U(1)$ Maxwell field is given by
\be\label{maxw}
F_{tr}=\frac{Q}{r},\label{metric3}
 \ee
where $Q$ is the
electric charge. Although these solutions for
$r\to\infty$ are asymptotically AdS,  
they have a power-law curvature singularity at $r=0$,
 where $R\sim \frac{8\pi GQ^2}{r^2}$. This $r\to 0$ behavior of the
 charged BTZ black hole has to be compared with that of the uncharged
 one, for which $r=0$ represents just a singularity of the causal 
 structure. For $r> l$, the charged black hole is described by the
Penrose diagram as usual \cite{Kogan:1992nh}.

In the present paper we will consider for simplicity only the non-rotating case
(i.e. we  will set $J=0$), however our results can be easily extended to
the charged, rotating BTZ black hole.

In the $J=0$ case the  black hole has two, one or no horizons, depending
on whether \be \label{ev}\Delta=8GM-4\pi G Q^2(1-2\ln \sqrt{\pi
G}\frac{2Ql}{l}) \ee is greater than, equal to or less than zero,
respectively. The Hawking temperature $T_H$ of the black hole
horizon is
\be \label{tem}T_H=\frac{r_+}{2\pi l^2}-\frac{2 G
Q^2}{r_+}.\ee
According to the Bekenstein-Hawking formula, the
thermodynamic entropy of a black hole is proportional to the area $A$ of
the outer event horizon,  $S=\frac{A}{4G}$. For the charged BTZ black hole we have
\be \label{en}
S=\frac{\pi r_+}{2G}= \frac{\pi l}{G} \sqrt{2GM+ 2\pi
GQ^{2}\ln\frac{r_{+}}{l}}.\ee

\section{Asymptotic symmetries}
It is a well-known fact that the asymptotic symmetry group (ASG) of
AdS$_{3}$, i.e. the group that leaves invariant the asymptotic form of
the metric,
is the conformal group in two spacetime dimensions \cite{Brown:1986nw}.
This fact  supports the  AdS$_{3}$/CFT$_{2}$ correspondence
\cite{Maldacena:1997re,Witten:1998qj} and has
been used to calculate the microscopical entropy of the BTZ black
hole \cite{Strominger:1997eq}. In order to determine the ASG   one has first  to fix boundary
conditions for the fields at $r=\infty$ then to find the Killing
vectors leaving these boundary conditions invariant.

The boundary conditions  must be relaxed enough to allow for the
action of the conformal group and for the right boundary
deformations, but tight enough to keep finite the charges associated
with the ASG generators, which are given by boundary terms of the
action (\ref{ac2}).
For the uncharged BTZ black hole suitable boundary conditions for the  
metric are
\cite{Brown:1986nw}
\ba\lb{bc}
g_{tt}&=& -\frac{r^{2}}{l^{2}}+\order(1),\quad g_{t\theta}= \order(1),\quad
g_{tr}=g_{r\th}= \order(\frac{1}{r^{3}}),\nonumber\\
g_{rr}&=& \frac {l^{2}}{r^{2}}+\order(\frac{1}{r^{4}}),\quad g_{\th\theta}=
r^{2}+\order(1),
\ea
whereas the vector fields preserving them are
\ba\lb{vf}
&&\chi^{t}=l \left(\e^{+}(x^{+})+\e^{-}(x^{-})\right)+
\frac{l^{3}}{2r^{2}}(\partial^{2}_{+}\e^{+}+\partial^{2}_{-}\e^{-})+
\order(\frac{1}{r^{4}}),\nonumber \\
&&\chi^{\theta}=\e^{+}(x^{+})-\e^{-}(x^{-})
-\frac{l^{2}}{2r^{2}}(\partial^{2}_{+}\e^{+}-\partial^{2}_{-}\e^{-})+
\order(\frac{1}{r^{4}}),\nonumber \\
&&\chi^{r}=-r (\partial_{+}\e^{+}+\partial_{-}\e^{-})+
\order(\frac{1}{r}),
\ea
where  $\e^{+}(x^{+})$ and $\e^{-}(x^{-})$ are arbitrary functions of
the light-cone coordinates $x^{\pm}= (t/l) \pm \th$ and
$\partial_{\pm}=\partial/\partial x^{\pm}$.
The generators $L_{n}$ ($\bar L_{n})$ of the diffeomorphisms with
$\e^{+}\neq 0$ ($\e^{-}\neq 0$) obey the Virasoro algebra

\ba\label{va}
&&[L_{m},L_{n}]=(m-n) L_{m+n}+ \frac{c}{12}(m^{3}-m)\delta_{m+n\,0}\nonumber\\
&&[{\bar L}_{m},{\bar L}_{n}]=(m-n){\bar L}_{m+n}+\frac{c}{12}(m^{3}-m)
\delta_{m+n\,0}\nonumber\\
&&[L_{m}, \bar L_{n}]=0,
\ea
where $c$ is the central charge. In the semiclassical regime $c>>1$, explicit 
computation of  $c$ gives
\cite{Brown:1986nw}
\be\lb{cc}
c=\frac{3l}{2G}.
\ee

The previous construction in principle should work for every 3D geometry which is
asymptotically AdS. However, it is not difficult to realize that it
works well only  for the uncharged  BTZ black
hole (\ref{metric}). In its implementation  to the charged
case one runs in two main problems. First, the boundary
conditions (\ref{bc}) do not allow for the  term in Eq. (\ref{charged}) 
describing boundary deformations behaving as $\ln r$.
One could relax the boundary conditions by allowing for  such terms, but
this will produce divergent boundary terms. Second, if the black hole is
charged we must also provide
boundary conditions at $r\to\infty$ for the electromagnetic field.
In view of Eq. (\ref{maxw}), simple-minded boundary conditions would
require $F_{tr}=Q/r + \order(1/r^{2}),\quad F_{t\th}=
\order(1/r),\quad F_{r\th}= \order(1/r)$. However, these boundary conditions
are not invariant under diffeomorphisms generated by the Killing
vectors (\ref{vf}). Again, one could relax the boundary conditions,
but then one should take care that the  associated boundary terms
 remain finite.
Both difficulties can be solved by relaxing the boundary conditions
for the metric and for the EM field and by using a suitable
renormalization procedure to keep the boundary terms finite.

In the coordinate system $(r,x^{+},x^{-})$  suitable
boundary conditions
for the metric, as $r\to\infty$, are
\ba\label{bc1}
g_{+-}&=& -\frac{r^{2}}{2} +\Gamma_{+-}\ln \frac{r}{l}+
\gamma_{+-}+\order(\frac{1}{r}),\nonumber\\
g_{\pm\pm}&=&  \Gamma_{\pm\pm}\ln \frac{r}{l}+
\gamma_{\pm\pm}+\order(\frac{1}{r}),\nonumber\\
g_{\pm r}&=& \Gamma_{+-}\frac{\ln \frac{r}{l}}{r^{3}}+
\frac{\gamma_{\pm r}}{r^{3}}+\order(\frac{1}{r^{4}¥}),\nonumber\\
g_{rr}&=& \frac{l^{2}¥}{r^{2}} +\Gamma_{rr}\frac{\ln \frac{r}{l}}{r^{4}}+
\frac{\gamma_{rr}}{r^{4}}+\order(\frac{1}{r^{6}¥}),
\ea
where the  fields
$\gamma_{\mu\nu}(x^{+},x^{-}),\,\Gamma_{\mu\nu}(x^{+},\,x^{-}),\,\mu,\nu=r,+,-$
are function of $x^{+},x^{-}$ only and describe deformations of the $r=\infty$
asymptotic conformal boundary of AdS$_{3}$.
One can easily check that the boundary conditions (\ref{bc1}) remain
invariant under the diffeomorphisms generated
by $\chi^{r}$  of Eq.
(\ref{vf}) and by the  other two Killing vectors, which in light-cone 
coordinates  take the form
\be\lb{vf1}
\chi^{\pm}=2\e^{\pm}+
\frac{l^{2}}{r^{2}}\partial^{2}_{\mp}\e^{\mp}+
\order(\frac{1}{r^{4}}).
\ee
The generators of ASG span the virasoro algebra (\ref{va}), and the
boundary fields $\gamma,\Gamma$ transform as 2D conformal field of
definite weight with (possible) anomalous terms.

A set of boundary conditions for the EM field $F_{\mu\nu}$ that are left invariant under the action
of the ASG  generated by $\chi^{\mu}$ are
\be\label{bc2}
F_{+-}=\order(1),\quad F_{+r}=\order(\frac{1}{r}),\quad F_{-r}=\order(\frac{1}{r}).
\ee
Notice that we are using very weak boundary conditions for the EM
field. We allow for deformations of the EM field which
are of the same order of the classical background solution (\ref{maxw}).
Although the boundary conditions are left invariant under the action
of the ASG the classical solution (\ref{maxw}) is not. Thus, we are
using  a broader notion of asymptotic symmetry, in which the
classical background solution for matter fields (but not that for the
gravitational field) may change under the  action of the ASG.
This  broader
notion of ASG  remain self-consistent because, as  we will see
 in detail in the next section,  the contribution of
the matter fields to the boundary terms generating the boundary
charges  vanishes.

\section{Boundary charges and statistical entropy}
In the previous section we have shown that, choosing suitable boundary
conditions, the  deformations of the charged BTZ black hole can be
generated by the action of the conformal group in 2D.
However, the weakening of the boundary conditions with respect to the
uncharged case is potentially dangerous,  because it can result in
divergences of the charges
associated with the generators of the conformal algebra.

In the case of the uncharged BTZ black hole, these charges can be
calculated using a canonical realization of the ASG
\cite{Brown:1986nw,NavarroSalas:1998ks,NavarroSalas:1999sr}.
Alternatively, one can use a lagrangian formalism and work
out  the stress energy tensor  for the boundary CFT
\cite{Balasubramanian:1999re}.  The relevant information we are interested in 
is
represented by the charge $l_{0}=\bar l_{0}$ associated with the
Virasoro operators $L_{0}$ and $\bar L_{0}$ (we are considering the
spinless case) and by the central charge $c$
appearing in algebra (\ref{va}).
 The information about $l_{0}$
and $c$  is encoded in the
boundary stress-energy tensor $\Theta_{\pm\pm}$ of the 2D CFT.
It can be calculated  either using the
Hamiltonian or the lagrangian formalism  and  expressed
in terms of 
the  fields describing boundary deformations.
For the uncharged BTZ black hole one finds
\be\label{bse}
\Theta_{\pm\pm}= \frac{1}{4 l G} \gamma_{\pm\pm},
\ee
where $\gamma_{\pm\pm}$ are the boundary fields parametrizing the
$\order(1)$ deformations in the $g_{\pm\pm}$ metric components.
Using the classical field equations one can show that 
$\gamma_{\pm\pm}$ are chiral functions, i.e. $\gamma_{++}$ 
($\gamma_{--}$) is function 
of $x^{+}$ ($x^{-}$) only \cite{NavarroSalas:1998ks}. 

Passing to consider the charged BTZ black hole, we have to worry both
about contribution to $\Theta_{\pm\pm}$ coming from the EM field and
about divergent terms originating from the  $\ln r$ terms
in Eq. (\ref{bc1}). From general grounds, the contribution of matter
fields are expected to fall off
 for  $r\to\infty$ more rapidly then those coming from the gravitational terms  and
from the cosmological constant.
Thus, as anticipated in the previous section, the EM
part of the action gives a vanishing contribution to $\Theta_{\pm\pm}$.
This can be explicitly shown by working out explicitly the surface term
$I^{(EM)}_{bound}¥$
one has to add to the action (\ref{ac2}) in order to make functional
derivatives with respect to the EM potential vector $A_{\mu}$ well
defined. One has
\be\label{e1}
\delta I^{(EM)}_{bound}\propto \int¥ d^{2}x \sqrt{-g^{(3)}}
N_{\mu}F^{\mu\nu}\delta A_{\nu},
\ee
where $N_{\mu}$ is a unit  vector normal to the boundary.
Using the boundary conditions (\ref{bc1}) and (\ref{bc2}) one finds
$\delta I^{(EM)}_{bound}= \order(1/r)$, giving a vanishing contribution
when the boundary is pushed to $r\to\infty$.
The same result can be reached considering the Hamiltonian. In this
case
variation of the EM part of the Hamiltonian
gives the boundary term
\be\label{e2}
\delta H^{(EM)}¥_{bound}\propto \int¥ d\th A_{t} \delta \pi^{r}
N^{r},
\ee
where $\pi^{r}$ denote the conjugate momenta to $A^{r}$.

Conversely, the $\ln r$ terms appearing in the asymptotic expansion (\ref{bc1})
give divergent contributions to the surface term. This fact has been
already noted in Ref. \cite{Martinez:1999qi}, where a renormalization procedure was
also proposed.  One encloses the system in a circle of radius $r_{0}$
and, in the limit $r\to \infty$, one takes also $r_{0}¥\to \infty$ keeping
the ratio $r/r_{0}=1$.
This renormalization procedure can be easily implemented to define a
renormalized black hole mass $M_{0}(r_{0})$, which has to be interpreted as the
total energy (electromagnetic and gravitational) inside the circle of
radius $r_{0}$. We have just to  write the metric function  (\ref {charged}) as
$f(r)= r^{2}/l^{2}-8GM_{0}(r_{0})-8\pi G Q^{2}\ln(r/r_{0})$ with
\be\label{mass}
M_{0}(r_{0})=M+\pi Q^{2}¥\ln(\frac{r_{0}}{l}).
\ee
Taking now the limit $r,r_{0}\to \infty$, keeping $r/r_{0}=1$, the
third term in $f(r)$ vanishes, leaving just the renormalized  mass term.
Moreover, because the total energy of the system cannot depend on
the value of $r_{0}$, we can take $r_{0}=r_{+}$, so that the total
energy is just $M_{0}(r_{+})$, the renormalized mass evaluated on the
outer horizon.

The same renormalization procedure can be  easily implemented for the
boundary deformations in Eq. (\ref{bc1}).
We just define renormalized deformations
\be\label{rdef}
\gamma^{(R)}_{\pm\pm}=\gamma_{\pm\pm}+ \Gamma_{\pm\pm}¥
\ln\frac{r_{0}}{l},
\ee
and similarly for $\gamma^{(R)}_{+-}$,$\gamma^{(R)}_{\pm
r}$,$\gamma^{(R)}_{rr}$, such that the boundary conditions (\ref{bc1})
become
\be\label{e7}
g_{\pm\pm}=
\gamma^{(R)}¥_{\pm\pm}+ \Gamma_{\pm\pm}\ln \frac{r}{r_{0}}+\order(\frac{1}{r}),
\ee
and similar expressions for $g_{+-}, g_{\pm r}, g_{rr}$.
In the limit $r,r_{0}\to \infty$, with $r/r_{0}=1$, the $\ln(r/r_{0})$ term in Eq.
(\ref{e7}) vanishes and we are left with boundary conditions which
have   exactly the same form of those for the uncharged BTZ
black hole but with the boundary fields $\gamma_{\mu\nu}$ replaced by
the renormalized boundary deformations (\ref{rdef}).
It follows immediately that the stress-energy tensor for the
boundary  CFT dual to the charged BTZ black hole is
\be\label{bse1}
\Theta_{\pm\pm}= \frac{1}{4 l G} \gamma^{(R)}¥_{\pm\pm},
\ee
with $\gamma^{(R)}_{\pm\pm}$ given by Eq. (\ref{rdef}). One can also check 
that the field equations (\ref{ein}) imply that 
$\gamma^{(R)}¥_{\pm\pm}$  have to be chiral functions of 
$x^{\pm}$, respectively.  

The central charge of the 2D CFT can be calculated using the
anomalous transformation law for $\gamma^{(R)}_{\pm\pm}$ under the
conformal transformations generated by (\ref{vf1}),
\be\label{ta}
\delta_{\e^{\pm}¥}\gamma^{(R)}_{\pm\pm}= 2(\e^{\pm}\partial_{\pm}+ 2
\partial_{\pm}\e^{\pm})\gamma^{(R)}_{\pm\pm}-
l^{2}\partial^{3}_{\pm}\e^{\pm}.
\ee
As expected, it turns out that the central charge is given by Eq. (\ref{cc}).
The charge associated to
time translations, $l_{0}=\bar l_{0}$,  can be calculated using Eq. (\ref{bse1}). One
obtains
\be\label{l0}
l_{0}= \frac{1}{2}l M_{0}(r_{+}),
\ee
where $M_{0}$ is the renormalized black hole mass (\ref{mass}).

In the semiclassical regime of large black hole mass, the existence
of an AdS$_{3}$/CFT$_{2}$ correspondence implies that the number of
excitations of the AdS$_{3}$ vacuum with mass $M$ and charge $Q$
should be counted by the asymptotic growth of the number of states 
in the CFT \cite{Cardy:1986ie},
\be\label{cardy}
S=4\pi\sqrt{\frac{cl_{0}}{6}}.
\ee
Using Eqs. (\ref{cc}), (\ref{l0}) and (\ref{mass}) we get

\be
S= 4\pi l\sqrt{\frac {M_{0}}{8G}}
=\frac{\pi l}{2G}\sqrt{8GM+8\pi
GQ^{2}\ln(\frac{r_{+}}{l})},
\ee
which matches exactly the Bekenstein-Hawking entropy of the charged
BTZ black hole (\ref{en}).

In this paper we have shown that the
Bekenstein-Hawking entropy of the charged BTZ black hole can be
exactly reproduced by counting states of the CFT generated by 
deformations of the AdS$_{3}$ boundary. The difficulties related  
with the presence of a curvature singularity have been circumvented 
using a renormalization procedure. Our result shows that  
the notion of asymptotic symmetry and  related machinery can be 
successfully used to give a microscopic meaning to the thermodynamical 
entropy of black holes also in the presence of curvature singularities.
In particular, this result could be very important for the 
generalization to the higher dimensional case of low-dimensional 
gravity  
methods for calculating the statistical entropy of black holes.

\end{document}